\documentclass{article}
\usepackage{spconf,amsmath,graphicx, amsfonts,,amsthm,amssymb,enumerate,enumitem,mathtools,float,accents,hyperref,url,enumerate, xcolor, makecell, subcaption}

\graphicspath{ {./figures/} }

\DeclarePairedDelimiter{\norm}{\lVert}{\rVert}

\title{PerMod: Perceptually Grounded Voice Modification with Latent Diffusion Models}
\makeatletter
\def\@name{
  \emph{Robin Netzorg}$^{1}$,
  \emph{Ajil Jalal}$^{1}$
  \emph{Luna McNulty}$^{2}$,
  \emph{Gopala Krishna Anumanchipalli}$^{1}$
  \thanks{This work was supported by the UC Noyce Initiative, Society of Hellman Fellows, NSF, NIH/NIDCD and the Schwab Innovation Fund.}
}
\makeatother
\address{
  $^1$University of California, Berkeley,
  $^2$Brown University
}
\copyrightnotice{979-8-3503-0689-7/23/\$31.00~\copyright2023 IEEE}
\begin{document}
%\ninept
%
\maketitle
\begin{abstract}
Perceptual modification of voice is an elusive goal. While non-experts can modify an image or sentence perceptually with available tools, it is not clear how to similarly modify speech along perceptual axes. Voice conversion does make it possible to convert one voice to another, but these modifications are handled by black box models, and the specifics of what perceptual qualities to modify and how to modify them are unclear. Towards allowing greater perceptual control over voice, we introduce PerMod, a conditional latent diffusion model that takes in an input voice and a perceptual qualities vector, and produces a voice with the matching perceptual qualities. Unlike prior work, PerMod generates a new voice corresponding to specific perceptual modifications. Evaluating perceptual quality vectors with RMSE from both human and predicted labels, we demonstrate that PerMod produces voices with the desired perceptual qualities for typical voices, but performs poorly on atypical voices.
\end{abstract}
\section{Introduction}
\label{sec:intro}

% Modifying speech is a non-trivial task. Unlike other data modalities where non-experts can use their perception to modify the data in question, like dragging the pose a creature in an image \cite{pan2023drag} or changing the words of a sentence to evoke a particular response, speech does not lend itself easily to modification. Although people can easily recognize certain qualities of voice, such as whether the voice is masculine or feminine, or whether or not the voice sounds happy or neutral, recognizing and describing the perceptual qualities of a voice that make it sound masculine or feminine, or strained or unstrained is difficult even for experts \cite{pqvd2020}. 

Every data modality can be represented in a variety of forms, but speech is unique in that so many of its representations require expert understanding to be interpreted or modified. With a data modality like vision, non-experts regularly apply pixel-level filters and use deep learning methods to modify and generate images as desired. Current methods in speech do not allow for similar modification or generation. Voice conversion and text-to-speech systems generate voice clips from particular speakers, such as converting a masculine voice to a specific feminine voice. While current methods do allow for manipulations like prosody and emotion control \cite{hu2022prosodybert, zhou2022emotion}, they do not allow for modification in a larger perceptual space.

In contrast with voice manipulation methods' lack of perceptual flexibility, the human voice is incredibly flexible. From musical training for vocalists to rehabilitations for those suffering from voice pathologies to voice feminization or masculinization for transgender individuals, human beings are capable of manipulating their voices in remarkable ways. In order to teach and describe these manipulations to students and patients, speech language pathologists and vocal teachers have developed several intuitive descriptions of the perceptual qualities of voice, like strain, breathiness, roughness, and weight. Clinicians and teachers both undergo training to be able to recognize these qualities, and trying to quantify these qualities can result in disagreement between experts \cite{pqvd2020}. That said, these qualities serve as a useful tool for describing the unique characteristics of every voice.

With this flexibility in mind, we introduce the Perceptual Modification LDM (PerMod), a latent diffusion model that is the first to allow for perceptual modification of a speaker's voice. Taking as input a speaker's voice and conditioning on a vector of 7 perceptual qualities, PerMod outputs a modified voice with the desired perceptual qualities, as measured by root mean square error (RMSE). We demonstrate that PerMod is capable of modifying typical voices, but that modification of atypical voices needs improvement.\footnote{A demo of PerMod is available at https://berkeley-speech-group.github.io/PerModLDM/.} We end with a discussion of the limitations of PerMod and future work. 

% \noindent Our contributions are as follows:

% \begin{itemize}
%     \item The addition of resonance and weight labels to the Perceptual Voice Qualities Database.

%     \item The combination of perceptual vocal qualities from multiple fields and condensing them into an 8-dimensional perceptual vector in order to measure the vocal quality of a voice.

%     \item The introduction of PerMod, the first model to perform perceptually-grounded voice modification over a space of multiple vocal qualities. 
% \end{itemize}

\begin{figure*}
    \centering
    \includegraphics[width=0.65\textwidth]{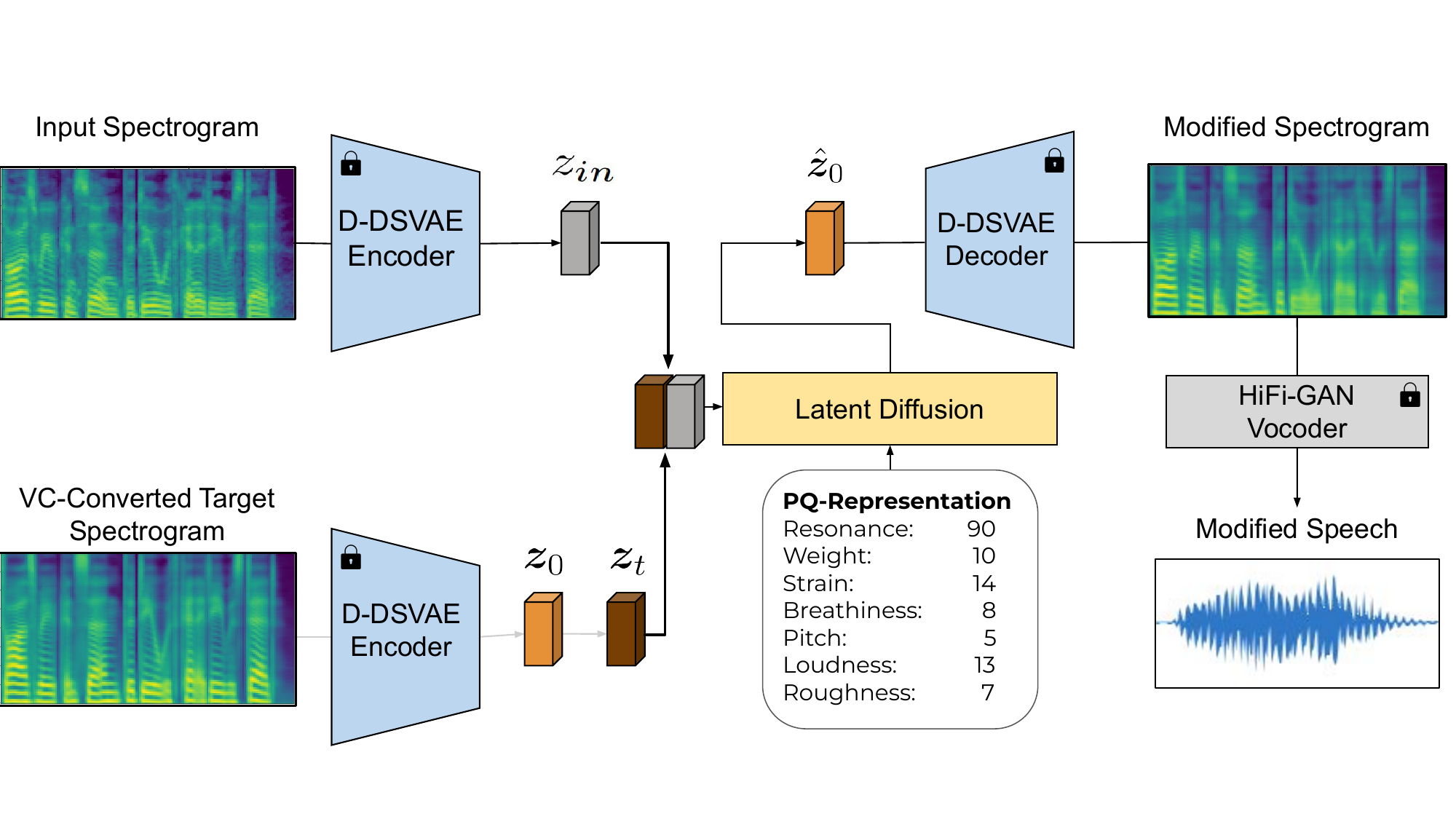}
    \caption{High-level overview of PerMod during training. Given an input $\textbf{x}_{in}$, a target $x_{out}$, and the target's PQ-Representation $c_{per}$, PerMod aims to learn to generate a new audio clip with the target's perceptual qualities. D-DSVAE weights are frozen during training.}
    \label{fig:system_overview}
\end{figure*}

\section{Related Work}
\label{sec:related}

\subsection{Diffusion Models} 
Diffusion models~\cite{ddpm, score, score_sde} have shown unprecedented success in multiple modalities, such as text-conditional image and audio generation~\cite{stable_diffusion, liu2023audioldm}, \emph{de novo} protein design~\cite{diffusion_proteins}, natural language processing~\cite{diffusion_nlp}, and medical imaging~\cite{jalal_mri, song_mri}. For an extensive survey on the current state of diffusion models and their applications, see~\cite{diffusion_survey}.

Within the context of perceptual modification of speech, diffusion models allow for great flexibility in choosing the perceptual features and their associated values. 

% \subsection{Audio and Speech Manipulation}

% While methods allow for prosody control \cite{hu2022prosodybert}, or emotion control \cite{zhou2022emotion}, no method currently allows for modification of voice by changing desired perceptual voice qualities. 

\subsection{Perceptual Voice Quality}

Thought of as the acoustic ``coloring'' of an individual's voice, voice quality has long been a topic of interest in speech language pathology and processing \cite{kreiman1993perceptual, kellervoicequality2005}. From the emotion to vocal fry and breathiness, vocal quality spans the different subjective perceptions of a voice. Speech processing is well aware of vocal quality, with prior work noting that voices with uncommon or pathological vocal qualities lead to poor performance for spoken language processing systems if not taken into consideration \cite{kellervoicequality2005, lee2016automatic}. 

The question of how voice quality can be used to identify and modify an individual's voice is crucial to our work. In speech language pathology (SLP), experts will use vocal quality to perform initial diagnosis of an individual's voice \cite{capev2009, pqvd2020}. In non-surgical treatment of voice, clinicians often have a patient perform vocal exercises to bring certain voice qualities into healthy levels \cite{capev2009}. Vocal quality sees much attention outside of SLP. Musical vocal teachers or voice feminization/masculinization teachers will use other vocal qualities, such as vocal fry, timbre/resonance, and weight, to guide students towards a desired voice \cite{diamant2021examining, carew2007effectiveness}. 

Describing how to manipulate a voice can be a difficult task, but a complete mapping of perceptual qualities allows for more fine-grained and descriptive control of voice synthesis. Prior work in controllable voice synthesis has primarily focused on prosody control \cite{hu2022prosodybert}, or concerns a single perceptual quality, like emotion \cite{zhou2022emotion}. Voice conversion methods will shift perceptual qualities, but often do so through end-to-end training and do not allow for fine-grained modification of a voice \cite{lian2022robust}. In this work, we aim to produce a voice modification system that considers multiple forms of perceptual qualities, and allows for direct control over them. 

\section{Perceptual Voice Qualities}

In this section, we describe the modeling of perceptual voice qualities. While the space of possible perceptual voice qualities is vast, we limit the scope of our perceptual qualities to seven perceptual features, as introduced in prior work \cite{netzorg2023interpretable}. Additionally, we describe how we model these perceptual qualities using the ComParE 2016 feature set \cite{weninger2013acoustics}. For the rest of this work, we will often refer to perceptual voice qualities as perceptual qualities (PQs), and the vector of perceptual qualities as the PQ-Representation.

\subsection{Perceptual Voice Qualities Database (PVQD)}

In clinical settings, perception assists greatly in the early stages of diagnosis of voice pathologies. Often times, clinicians will use rating scales like the Consensus Auditory-Perceptual Evaluation of Voice (CAPE-V) to provide early information on possible voice pathologies an individual may have \cite{capev2009}. Clinicians undergo training to accurately rate these perceptual voice qualities \cite{pqvd2020}.

To integrate expert labeling into our system, we utilize the Perceptual Voice Qualities Database (PVQD), which serves a publicly available ratings of perceptual voice qualities from the CAPE-V scale \cite{pqvd2020}. The PVQD includes 296 audio files of around 30 seconds of audio, whereby a speaker follows the CAPE-V evaluation protocol, and reads six sentences and produces vowels $/a/$ and $/i/$ for 1-2 seconds. The authors behind the PVQD had each audio clip rated by three separate clinicians across two trials according to the CAPE-V scale. 

\subsection{Gendered Perceptual Qualities}

While the PVQD serves its purposes as a diagnostic tool, for voice generation and modification, it is incomplete. General measures of perceptual voice qualities, such as the timbre or vocal mass, are missing. Attempting to perform manipulations that are common in voice conversion, such as converting a masculine voice to a feminine voice, would not be possible with the  CAPE-V scale's ratings of deviation alone. As such, we utilize an expanded form of PVQD, PVQD+, which includes labels of resonance and weight that are provided by a voice teacher who specializes in transgender voice training \cite{netzorg2023interpretable}. Vocal resonance and vocal weight correspond to the two main physiological differences that distinguish masculine and feminine voices voices \cite{markova2016age}. Physically, resonance corresponds to the amount of space above the vocal folds in the vocal tract. More space causes lower resonant frequencies to be amplified \cite{kent1993vocal}, resulting in perceptually deeper timbre, even at high pitches. Weight corresponds to the vibratory mass of the vocal folds, which is correlated with the open quotient, (the proportion of the glottal cycle during which the vocal folds are open) and spectral slope (the decline in amplitude from the first to the Nth harmonic) \cite{zhang2016cause}. These ideas are also often used in singing lessons for vocalists, but the focus on voice modification along perceptions of gender is more directly applicable to voice modification via a neural network.

% Such lessons often involve individuals changing the resonance (timbre resulting from the amount of space in the vocal tract) and weight (vocal mass) to match those of the patient's assigned gender. 

In the PVQD+, the voice teacher listened to the entirely of PVQD and provided a label of resonance and weight on a scale 1-100, similar to the CAPE-V scale. For resonance, a value of 1 represented the darkest resonance possible and a value of 100 represented the brightest resonance possible. Similarly for weight, a value of 1 represented the lightest voice possible and a value of 100 represented the heaviest voice possible. Typical feminine voices were given a resonance value of 90 and a weight value of 10, and the opposite for typical masculine voices. With these labels, we expand the number of usable samples in underrepresented voices. We provide a summary of the final perceptual qualities in Table~\ref{tab:pvq}.

\begin{table}[]
    \centering
    \begin{tabular}{|c|c|c|}
        \hline
         \textbf{PQ} & \textbf{Type} & \textbf{Description}  \\
        \hline
         Resonance & Gendered & \makecell{Sound quality of the \\size of the vocal tract} \\
         \hline
         Weight & Gendered & \makecell{Sound quality of the\\ vocal fold vibratory mass} \\
         \hline
         Strain & CAPE-V & \makecell{Perception of excessive\\ vocal effort (hyperfunction)} \\
         \hline
         Loudness & CAPE-V &Deviation in loudness \\
         \hline
         Roughness & CAPE-V & \makecell{Perceived irregularity \\in the voicing source} \\
         \hline
         Breathiness & CAPE-V &  Audible air escape in the voice \\
         \hline
         Pitch & CAPE-V & Deviation in pitch\\

        \hline
    \end{tabular}
    \caption{The complete list of perceptual qualities used in PerMod and their descriptions. Resonance and weight are taken from vocal training, and the other qualities are taken from the CAPE-V protocol.}
    \label{tab:pvq}
\end{table}

\subsection{Perceptual Modeling}\label{rf_model}

 While recent work has demonstrated that non-experts can label perceptual qualities with remarkable accuracy \cite{netzorg2023interpretable, mcallisterperceptual2023}, the perceptual labeling of an entire dataset needed for pretraining a usable voice synthesis model is a prohibitive task. As such, we instead model the perceptual qualities of PVQD, and apply the learned model to VCTK for pretraining. We achieve this by training a Random Forest (RF)  on a train-validation-test split (60-20-20) of PVQD. Using absolute deviation as the evaluation metric, the random forest achieves an  $12.34$ test RMSE averaged across perceptual qualities. 

 We note that in PVQD, the perceptual label is assigned to a 30 second clip that the clinician or voice teacher has considered holistically. When applying the random forest to VCTK to generate perceptual quality vectors, it is applied at a sentence-level per utterance on a roughly 3-5 second clip of audio. We hypothesize that additional sentence-level labels for VCTK from clinicians and voice teachers would improve the performance of PerMod, but we leave this to future work.

\section{Perceptual Modification LDM}
\label{sec:model}

In this section, we describe the Perceptual Modification LDM (PerMod), which consists of a conditional latent diffusion model that, given a PQ-Representation, modifies an input voice and returns an output voice with the perceptual qualities of the conditional PQ-Representation. For a visual overview of the system, refer to Figure \ref{fig:system_overview}.

\subsection{System Overview}

\subsubsection{D-DSVAE Encoder/Decoder}

The key goal of this work is to modify the speaker information independently from the content information. As such, we use the Denoising Disentangled Sequential VAE (D-DSVAE) \cite{lian2022robust}, which simultaneously embeds a given mel-spectrogram into a speaker and content embedding. The D-DSVAE is training procedure is similar to that of a regular sequential VAE, but, in addition to a reconstruction loss $\mathcal{L}_{REC}$, splits the KL-Divergence term into $\mathcal{L}_{KLD_S}$ and $\mathcal{L}_{KLD_C}$, which trade-offs between time-invariant information (speaker) and time-variant information (content). The trade-off between the two KL terms is controlled via hyperparameters $\alpha, \beta$ that are tuned during D-DSVAE training. The complete D-DSVAE loss is $\mathcal{L}_{\text{D-DSVAE}} = \mathcal{L}_{REC} + \alpha\mathcal{L}_{KLD_S} + \beta\mathcal{L}_{KLD_C}$.

One can perform voice conversion by switching the speaker embedding of one mel-spectrogram with that of another repeatedly over small enough time-scales, a trait we take advantage of during training (Section \ref{arch:paired_data}). 

\subsubsection{Perceptual Quality Representation}

As discussed in Section \ref{rf_model}, we model the perceptual qualities of a given audio segment via a random forest regressor trained on PVQD labels. This model generates a PQ-Representation, a $7$-dimensional vector $c_{per}$ that consists of the perceptual quality information of a given speech clip.

\subsubsection{Conditional Latent Diffusion Model}

PerMod is a conditional latent diffusion model, inspired by the success of similar models on audio editing tasks \cite{wang2023audit}. The model takes in as input $\mathbf{z}_{in}$, the D-DSVAE embedding of mel-spectrogram input $\mathbf{x}_{in}$, and aims to learn the distribution $p(\mathbf{z}_{out} | \mathbf{z}_{in}, c_{per})$, where $\mathbf{z}_{out}$ is the D-DSVAE embedding of the mel-spectrogram of target $\mathbf{x}_{out}$, and $c_{per}$ is the inferred perceptual information of $\mathbf{x}_{out}$. Taking $\mathbf{z}_{in}$ and $c_{per}$ as inputs, randomly selecting a timestep $t$ and generating noise $\mathbf{\epsilon}$, we produce $\mathbf{z}_t$, a noisy version of $\mathbf{z}_{out}$, via a noise scheduler. The diffusion model $\epsilon_{\theta}(\mathbf{z}_t, t, \mathbf{z}_{in}, c_{per})$ then generates sampled noise. The reconstruction loss between Gaussian noise $\mathbf{\epsilon}$ and the sampled noise $\epsilon_{\theta}(\mathbf{z}_t, t, \mathbf{z}_{in}, c_{per})$ guides training: 

\begin{equation}
    \mathcal{L}_{LDM} =
    \mathbb{E}_{(\mathbf{z}_{in}, \mathbf{z}_{out}, c_{per})}\mathbb{E}_{\mathbf{\epsilon}}\mathbb{E}_t \norm{\epsilon_{\theta}(\mathbf{z}_t, t, \mathbf{z}_{in}, c_{per})- \mathbf{\epsilon}}_2
\end{equation}

The architecture for the conditional LDM is a U-Net, which is the same architecture used in standard diffusion models and similar LDMs \cite{diffusion_survey}. To condition on the input audio, we concatenate $\mathbf{z}_{in}$ and $\mathbf{z_t}$, which means that the U-Net input has a channel length double that of the output channel length, producing $\mathbf{\hat{z}}_{out}$. Decoding $\mathbf{\hat{z}}_{out}$ with D-DSVAE, we produce the generated mel-spectogram $\mathbf{\hat{x}}_{out}$.

\subsubsection{Vocoder}

For the Vocoder, we utilize HiFi-GAN, which is frequently applied for high quality speech synthesis \cite{kong2020hifigan}, and is also the default decoder for D-DSVAE \cite{lian2022robust}. To adjust for different sampling rates, we use a HiFi-GAN that has been retrained on audio sampled at 16KHz. 

\subsection{Generating Paired Pre-Training Data}\label{arch:paired_data}

Given that the only information we are providing the diffusion model consists of an input audio clip and desired perceptual qualities, the training of this model requires that the content of $\mathbf{x}_{in}$ and $\mathbf{x}_{out}$ is consistent. VCTK, however, is not a paired dataset. To create an artificially paired dataset, we apply voice conversion with D-DSVAE to each sample $\mathbf{x}_{out}$ as preprocessing during training time. The resulting speech and embedding, $\mathbf{\hat{x}}_{match}, \mathbf{z}_{match}$, contain the speaker information of $\mathbf{x}_{out}$ and the content information of $\mathbf{x}_{in}$. We note that voice conversion is not perfect, and sometimes results in the generation of artifacts or unwanted noise. Fortunately, PerMod often treats these artefacts robustly, learning to filter them out at inference time.

% FastSpeech does something similar \cite{ren2019fastspeech}. 

\subsection{Fine-Tuning}

% Describe the naive fine-tuning set up. We leave applying Lora to future work.

Hoping to improve the ability of PerMod to produce atypical voices, we introduce a fine-tuning step on PVQD with Low-Rank Adaptation (LoRA), which allows for fast fine-tuning of larger pretrained models via rank-decomposition matrices  \cite{hu2021lora}. To make the data similar to that of VCTK, four of the six CAPE-V sentences are extracted from each audio sample in PVQD, with the average expert rating providing the PQ-Representation. As PVQD consists of speakers with both typical and atypical voices, all of whom read the same sentences, training on PVQD serves as a natural paired task that provides the opportunity to learn atypical-to-atypical modification (A2A), atypical-to-typical modification (A2T), and vice-versa (T2A).  

\subsection{Training}

We pretrain PerMod on a single A5000 GPU for 100 epochs over VCTK, which lasts roughly 24 hours. As LoRA allows for quick updates, we finetune on PVQD for 2000 epochs, which takes roughly 1 hour. 

\section{Experiments}
\label{sec:experiments}

Here, we report the results of pretrainining and fine-tuning PerMod, and the results of training a version of PerMod that only conditions on gendered PQs, named PerMod-RW. 

\begin{table*}[t]
    \centering
    \begin{tabular}{|c|c|c|c|c|c|c|c|}
        \hline
         \textbf{Model} & \textbf{Resonance} & \textbf{Weight} & \textbf{Breathiness} & \textbf{Loudness} & \textbf{Pitch} & \textbf{Roughness} &  \textbf{Strain} \\
        \hline
         PerMod-RW & 21.18 & 22.14 & - & - & - & - & - \\
         \hline
         PerMod-Pretrained & 16.93 & 19.00 & 31.22 & 26.70 & 30.13 & 28.53  & 27.35 \\
         \hline
    \end{tabular}
    \caption{Average RMSE of perturbed perceptual qualities vector and the perceptual qualities vector of the generated audio clip, using VCTK test files as input. The perceptual qualities vector is modified by changing an individual index to be a value between 10-90 with a step-size of 10.  Standard Error $<2.4$ for gendered perceptual qualities, and $<3.54$ for CAPE-V perceptual qualities.}
    \label{tab:eps_experiment}
\end{table*}

\subsection{Audio Quality}

In order to measure the audio quality of the three models, we use the Mean-Opinion-Score (MOS) of train and test clips from VCTK for PerMod-Pretrained and PerMod-RW, and the MOS of the D-DSVAE voice conversion as a baseline. We compute the MOS through Amazon Mechanical Turk (AMT), asking 5 workers to rate ~150 train and test clips post-modification, where 50 random audio clips from VCTK are passed into PerMod with the perceptual quality vectors of 3 train or test speakers. The results are summarized in Table~\ref{tab:audio_quality}. 

\begin{table}[]
    \centering
    \begin{tabular}{|c|c|c|}
        \hline
         \textbf{Model} & \textbf{MOS (Train)} & \textbf{MOS (Test)} \\
         \hline
         D-DSVAE VC  & 3.42 (+/- 0.09) & 3.32 (+/- 0.07)\\
         \hline
         PerMod-RW  & 3.48 (+/- 0.06) & 3.33 (+/- 0.05) \\
         \hline
         PerMod-Pretrained  & 3.49 (+/- 0.05) & 3.45 (+/- 0.05)\\
         \hline
         % PerMod-Finetuned  & 3.60 (+/- 0.05) &  3.45 (+/-0.06)\\
    \end{tabular}
    \caption{Average Mean-Opinion-Score (MOS) of the three PerMod models and D-DSVAE voice conversion on VCTK. A score of 1 is bad quality, and a score of 5 is excellent quality.}
    \label{tab:audio_quality}
\end{table}

We see that, for the training data, PerMod actually improves the overall quality of the generated audio when compared to the D-DSVAE voice converted audio, within a standard error for all models. On the test data, all models see a drop in performance, with PerMod-RW performing similar to D-DSVAE. Overall, we see that voice modification with PerMod maintains or improves the audio quality of voice conversion with D-DSVAE. 

\subsection{Typical-to-Typical (T2T) Modification}\label{sec:nat_eval}

To evaluate the ability of PerMod to produce voices with the desired perceptual qualities, we first report the RMSE over the two PerMod models, PerMod-RW and PerMod-Pretrained. Here, the RMSE is taken between the target's PQ-Representation and the generated audio's predicted PQ-Representation. The audio clips used to generate the train and test RMSE are the same as the audio clips used to compute the MOS. We report the results in Table~\ref{tab:rmse}.

For all the models, we see that the RMSE on both the train and test data is remarkably low. The train RMSE of PerMod-RW and PerMod-Pretrained are both within ranges of human error, with PerMod-Pretrained seeing higher test error. A possible explanation for the worse performance of PerMod-Pretrained is that, while the manifold of the perceptual qualities of the VCTK speakers is relatively uniform and the main source of variation between voices comes from the gendered perceptual qualities, PerMod-Pretrained lacks information on atypical voices to effectively modify the space of CAPE-V PQs. The CAPE-V perceptual qualities of voice are primarily measures of deviance, and the VCTK voices are all typical and fall within normal ranges for those perceptual qualities. PerMod-Pretrained does have a higher MOS test performance, implying that CAPE-V perceptual qualities, even for typical voices, are helpful for audio generation.

\begin{table}[]
    \centering
    \setlength{\tabcolsep}{4pt} % Adjust the padding between columns
    \begin{tabular}{|c|c|c|}
        \hline
         \textbf{Model} & \textbf{RMSE (Train)} & \textbf{RMSE (Test)} \\
         \hline
         
         PerMod-RW  & 6.32 (+/- 0.54) & 11.33 (+/- 1.12)\\
         \hline
         PerMod-Pretrained  & 11.06 (+/- 0.06) & 28.97 (+/- 2.93) \\
         \hline
         % PerMod-Finetuned  & 0.985 (+/- 0.002) & 0.993 (+/- 0.001)\\
         % \hline
    \end{tabular}
    \caption{Average RMSE across PQs between the target audio's perceptual quality and generated audio's predicted perceptual quality on VCTK.}
    \label{tab:rmse}
\end{table}

\subsection{Measuring Disentanglement of Perceptual Features}

The goal of PerMod is not to generate a particular voice, but generate voices with arbitrary vocal qualities. To measure how well PerMod achieves this, we perturb the PQ-Representation of a given voice. Given an audio clip and its associated PQ-Representation, we change one axis of the representation at a time with values between 10-90 with a step-size of 10, and then generate 10 audio clips with PerMod. We run the generation using 10 VCTK test clips as input, and report average RMSE across PQs in Table~\ref{tab:eps_experiment}. 

As seen in Table~\ref{tab:eps_experiment}, despite perturbing the perceptual qualities vectors, PerMod is able to generate, on average, audio clips with average test RMSE similar to the test RMSE when given a PQ-Representation that is associated with an actual voice. Similar to before, PerMod demonstrates higher levels of error on CAPE-V PQs, but, interestingly, the inclusion of CAPE-V information lowers the error for gendered PQs. Not shown in Table~\ref{tab:eps_experiment}, PerMod does see worse RMSE on the boundary of the perceptual qualities space, sometimes reaching higher RMSE when the perturbation is $10$ or $90$ for a given PQ. As these experiments show, PerMod can achieve similar levels of error on perturbed PQ-Representations, continuing to display a higher level of performance on gendered PQs, and worse performance on CAPE-V PQs.

\begin{figure*}[t]
     \centering
     \begin{subfigure}[b]{0.25\textwidth}
         \centering
         \includegraphics[width=\textwidth]{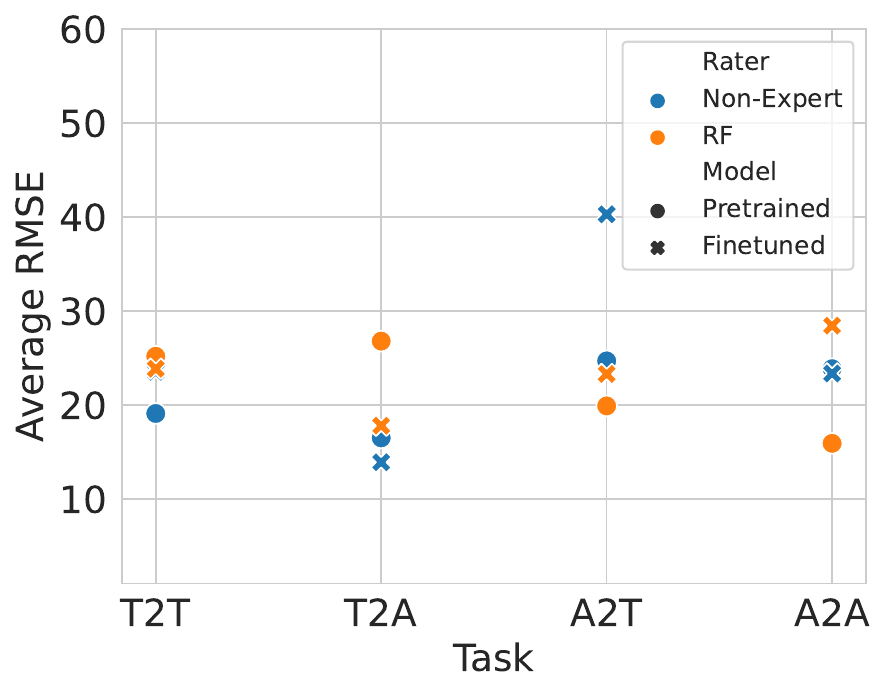}
         \caption{All PQs}
         \label{fig:avg_rmse}
     \end{subfigure}
     \begin{subfigure}[b]{0.25\textwidth}
         \centering
         \includegraphics[width=\textwidth]{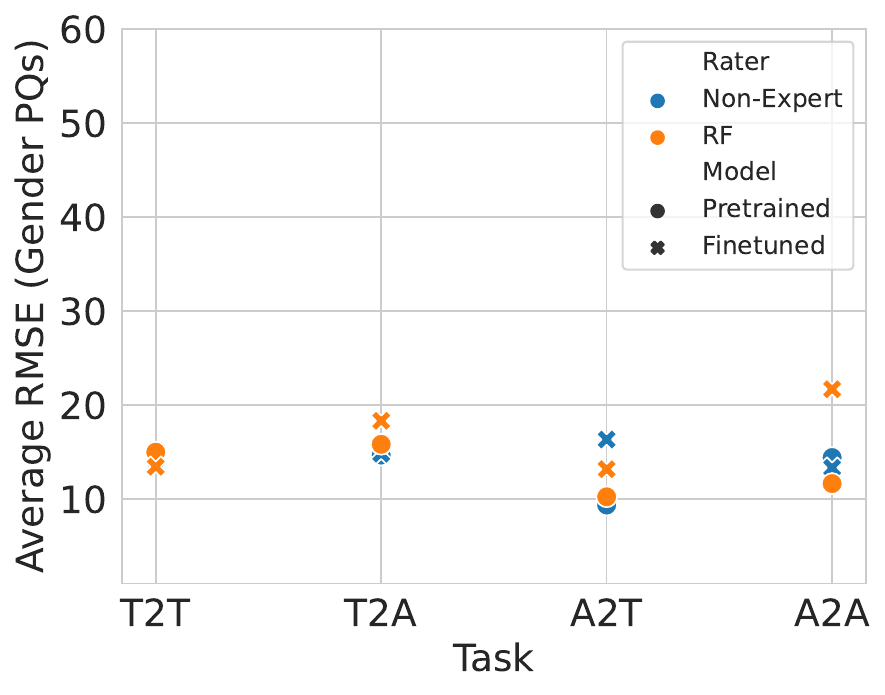}
         \caption{Gendered PQs}
         \label{fig:rw_err}
     \end{subfigure}
     \begin{subfigure}[b]{0.25\textwidth}
         \centering
         \includegraphics[width=\textwidth]{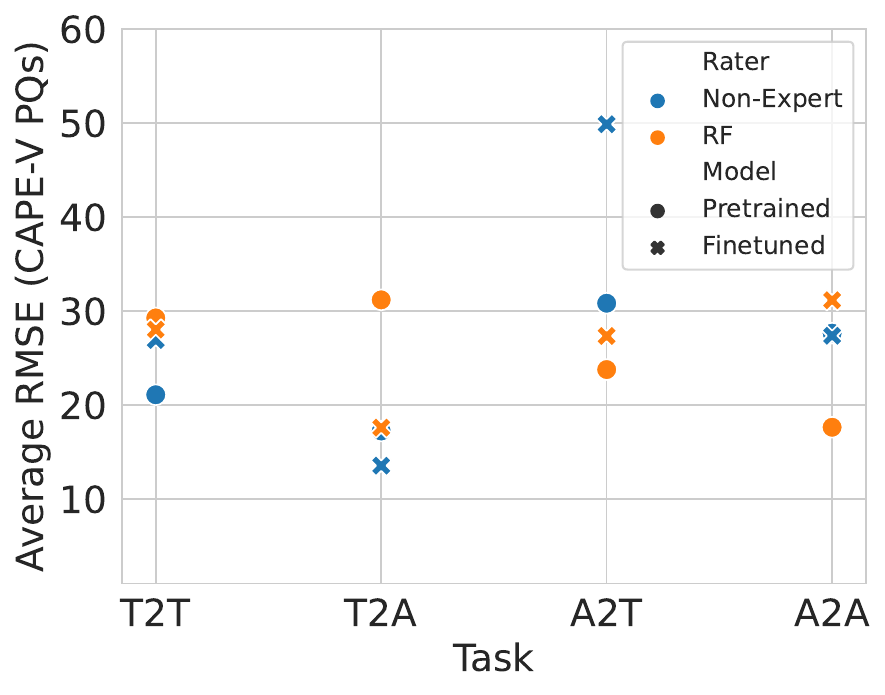}
         \caption{CAPE-V PQs}
         \label{fig:capev_err}
     \end{subfigure}

        \caption{Average RMSE three sets of PQs (All, Gendered, CAPE-V) for PerMod-Pretrained and PerMod-Finetuned, as measured by both average non-expert and predicted ratings. }
        \label{fig:errors}
\end{figure*}

\subsection{Modification of Atypical Speech (A2A, A2T, T2A)}

Up to this point, we have only explored the ability of PerMod to modify an input typical voice to an output voice with the typical perceptual vocal qualities, and we have only utilized a predictive model of perceptual quality as an evaluation metric of how well PerMod modifies perceptual qualities. Ideally, PerMod would be able to take as input a voice with \emph{any} PQ-Representation, and output a voice with those intended perceptual qualities, as perceived by the human ear. To this end, here we explore the modification of typical and atypical speech for the PerMod-Pretrained and PerMod-Finetuned models, including both evaluation from both the RF model of perceptual qualities and non-expert listeners.

As recent work has demonstrated, non-experts are capable of labeling perceptual qualities when given proper examples and instructions \cite{netzorg2023interpretable}. Non-experts do have a bias towards labeling atypical voices as more atypical than they actually are, resulting in an ensemble error roughly twice that of experts with an average RMSE across PQs of 23.45, but ensembles of non-experts nearly always label voices high in a particular PQ as such. Using AMT, we have six workers rate generated audio clips from both the pretrained and finetuned models, where input and target audio clips are randomly sampled from highly-atypical voices in PVQD and VCTK. This results in four voice modification tasks: Typical-to-Typical (T2T), Atypical-to-Typical (A2T), Typical-to-Atypical (T2A), and Atypical-to-Atypical (A2A). Finally, given a generated audio clip, the intended value for a particular PQ, and an average of non-expert ratings for that PQ, we calculate the RMSE across generated audio clips between the average non-expert rating and the intended PQ value. 

\begin{table}[]
    \centering
    \begin{tabular}{|c|c|c|}
        \hline
         \textbf{Task} & \textbf{PerMod-Pretrained} & \textbf{PerMod-Finetuned} \\
         \hline
         A2A & 23.87 & \textbf{23.37} \\
         \hline
         T2A & 16.51 & \textbf{13.92} \\
         \hline
         A2T & \textbf{24.70} & 40.30 \\
         \hline
         T2T & \textbf{19.11} & 23.55 \\
         \hline
    \end{tabular}
    \caption{Average RMSE across perceptual qualities between
Average Non-Expert PQ label and the target PQ value.}
    \label{tab:model_error}
\end{table}

Reporting the average RMSE across PQs across tasks in Table~\ref{tab:model_error}, several interesting findings emerge. Except for the performance PerMod-Finetuned on A2T, the highest RMSE is 24.70, and the lowest is 19.11, demonstrating that, on average, PerMod is able to generate audio clips that fall within non-expert error levels. While PerMod-Finetuned performed better on atypical modification tasks and PerMod-Pretrained performed better on typical modification tasks, except again for A2T, the differences in means are not statistically significant. As it stands, fine-tuning with LoRA does not seem to vastly improve the capability of PerMod to generate atypical voices, with the base pretrained model already achieving a similar performance.

Hoping to further understand the performance of PerMod-Pretrained and PerMod-Finetuned, we visualize the average RMSEs of three PQ groupings (All, Gendered, and CAPE-V) when compared with average non-expert labels and predicted random forest labels in Figure \ref{fig:errors}. Examining the performance on the different PQ groupings further outlines the capabilities and limits of PerMod. Regarding the gendered PQs of resonance and weight, across both non-expert and predicted labels, PerMod is able to produce voices with reasonable levels of error. With CAPE-V PQs, however, PerMod continues to perform worse, reaching higher levels of error as measured by non-experts and the RF model. PerMod struggles to modify atypical voices, but does see higher performance when making typical voices more atypical. Overall, however, PerMod struggles to modify voices along the CAPE-V PQs, revealing a current limitation of the system.

% \begin{figure}
%     \centering
%     \includegraphics[width=0.33\textwidth]{figures/RMSE_vs_MOS.pdf}
%     \caption{Average RMSE for both PerMod-Pretrained and PerMod-Finetuned plotted against Mean Opinion Score (MOS). }
%     \label{fig:rmse_vs_mos}
% \end{figure}

\section{Limitations and Future Work}
\label{sec:disc}

While PerMod is able to generate certain voices with desired perceptual qualities, on average, there is much room for improvement. Currently, PerMod underperforms on atypical voices, such as those in the PVQD. When given an atypical voice as input, PerMod-Pretrained and PerMod-Finetuned will often generate voices with lower audio quality. In the case of PerMod-Pretrained, the same inputs can also result in generating silence. The current fine-tuning approach, coupled with the lack of data, is not sufficient to produce voices with atypical values. In future work, we hope to expand PerMod with a richer dataset of atypical voices, and explore other fine-tuning or conditional generation approaches.

We hypothesize that there is a straightforward way to improve the generalizability of PerMod, without having to heavily modify the architecture: expanding the amount of labeled data available, particularly the amount of data on atypical voices. As prior work has demonstrated that non-experts can label perceptual qualities reasonably well \cite{netzorg2023interpretable, mcallisterperceptual2023}, we expect that increasing the number of human-labeled data samples would improve the ability of PerMod to generalize to atypical voices. 

\section{Conclusion}
\label{sec:conc}

In this work, we introduce the Perceptual Modification LDM (PerMod), a conditional latent diffusion model capable of modifying typical voices along perceptual lines. Utilizing insights from speech language pathology and transgender voice training, we incorporate perceptual voice qualities into a voice conversion pipeline, allowing for the modification of a given voice. Although PerMod is not currently able to modify atypical voices effectively, PerMod is a first step towards perceptually-grounded voice modification systems, which would allow for minor to major adjustments to an individual's voice.

\bibliographystyle{IEEEbib}
\bibliography{main}

\end{document}